# Many binaries among NEAs

D. Polishook [a,b,*] and N. Brosch [b]

[a] *Department of Geophysics and Planetary Sciences, Tel-Aviv University, Israel.*
[b] *The Wise Observatory and the School of Physics and Astronomy, Tel-Aviv University, Israel.*
[*] *Corresponding author: David Polishook, david@wise.tau.ac.il*

The number of binary asteroids in the near-Earth region might be significantly higher than expected. While Bottke and Melosh (1996) suggested that about 15% of the NEAs are binaries, as indicated from the frequency of double craters, and Pravec and Harris (2000) suggested that half of the fast-rotating NEAs are binaries, our recent study of Aten NEA lightcurves shows that the fraction of binary NEAs might be even higher than 50%.

We found two asteroids with asynchronous binary characteristics such as two additive periods and fast rotation of the primary fragment. We also identified three asteroids with synchronous binary characteristics such as amplitude higher than one magnitude, U-shaped lightcurve maxima and V-shaped lightcurve minima. These five binaries were detected out of a sample of eight asteroids observed, implying a 63% binarity frequency. Confirmation of this high binary population requires the study of a larger representative sample. However, any mitigation program that requires the deflection or demise of a potential impactor will have to factor in the possibility that the target is a binary or multiple asteroid system.

## Introduction

NEAs mitigation techniques are suggested with reference to objects size, orbit, composition and structure, assuming single objects. In this paper we point out the existence of a population of binary asteroids among NEAs which, according to our research, might be significantly higher than expected.

*Binary Asteroids*

The mere existence of binary asteroids was in doubt until 1993 when the spaceship Galileo obtained a close-up image of (243) *Ida* and *Dactyl*. Since then, binary asteroids have been found using a range of techniques: *i)* direct imaging using adaptive optics and large-aperture telescopes or the HST, can discover big Main Belt (MB) binaries with a large separation between the components, such as (45) *Eugenia* and (90) *Antiope*; *ii)* radar observations can detect much smaller binaries and satellites but this can be done only for near-Earth asteroids such as *1999 KW$_4$* and *2000 DP$_{107}$*; *iii)* time-series photometry can exhibit two additive periods in the asteroid lightcurve that reflect the rotational period of the primary body and the orbital period of the secondary fragment. (See review paper by Merline et al. 2002) This technique is best used for binaries with asynchronous periods where the different frequencies are easily noticed. Eclipses and occultations can also sometimes be seen in such lightcurves and can reveal binarity for synchronous asteroids even when the orbital and rotational periods are equal, such as for Pluto and Charon or for the NEA (69230) *Hermes*. In these cases, very steep V-shaped minima represent eclipse events while U-shaped maxima are produced by both components.

Up to June 2006, 96 binaries are known among the small bodies in the Solar System and one-third of them are NEAs – this is a very high fraction compared to the ~1% ratio of all known NEAs to all known MBAs. This anomaly can be solved by either of two explanations: a selection effect, with present-day techniques unable to detect binary MBAs with the characteristics of binary NEAs (e.g., small size, small separation, etc.), or a specific formation mechanism with the creation and evolution of binary asteroids depending on their Solar System location. Some ideas suggested for this dependence are collisions in the Main Belt, which can disrupt an asteroid into two or more fragments. The problem with such a scenario is that the disrupted fragment must not gain too much energy in order to remain in orbit around the main body. Another option, more likely to happen in the denser Main Belt, is the gravitational capture of a fragment from an asteroid by another. This scenario requires the presence of a third body in order to conserve angular momentum and energy. Bottke and Melosh (1996) suggested that when an NEA passes near a terrestrial planet it can be disrupted by the planet's tidal force. A fourth mechanism depends on the efficiency of re-emitting Sunlight. This effect, called YORP, is a torque on a rotating irregular body due to the absorption and subsequent re-emission of sunlight. When YORP spins up the asteroid (though spin down is another possibility) the body can pass the critical rotation speed threshold of a "rubble pile" structure and be disrupted into a binary (Bottke et al., 2002).

Pravec et al. (2006) recently summarized some important properties of asynchronous binary NEAs. The diameter ratios of secondaries to primaries (Ds/Dp) are almost always smaller than 0.5 when the primary diameter is less than two km. The secondaries are generally more elongated than the primaries, and orbit them in an almost circular orbit ($e<0.1$) with a semi-major axis that is 1.5-3 times the primary diameter. The secondaries' rotation periods are usually synchronized with the orbital period, both with a median value of ~16 hours with a distribution tail reaching ~11 hours on one side and at ~40 hours on the other. The distribution of the primaries' rotation period is more localized and ranges between 2.2 to 2.8 hours. This high spin rate is very close to the critical rate at which rubble pile asteroids disrupt, suggesting that the origin of such systems involved a fast rotating progenitor.

Known MBA and TNO binaries with asynchronous rotation have different properties, due to selection effects. Ds/Dp ranges from one (equal-sized components) to small values (tiny satellites). The distances between the primary and the secondary are larger than those of NEAs (otherwise they would not have been detected). The rotation periods of primaries are longer (several to a few tens of hours) and so are the orbital periods. The synchronous binaries' properties are similar to the asynchronous Main Belt binaries, including one NEA binary with synchronous motion, (69230) *Hermes*, that has a rotation period of 13.89 hours.

*The Aten Family*

Our study concentrated on the Aten family of the NEA group. The Atens are hard to observe from ground-based observatories for a major part of their orbits. As a result, the Atens properties derived from photometric observations, lightcurves, phase curves, color and spectroscopy, are not so well known in comparison to those of members of the Apollo and Amor families. Since Atens, by definition, cross the Earth's orbit, as do Apollo asteroids but not Amors, understanding their physical nature, composition and structure is essential for future mitigation attempts.

**Research**

We performed photometric observations of eight Atens to derive their rotation spin, triaxial shape, taxonomy, and phase curve parameters. The study is fully described in a recent paper (reference: Polishook and Brosch, 2006; submitted).

Only a few Atens could be observed in the survey that took place from mid-February to late May 2004 and from early April to mid-July 2005. Almost all Atens that could be, were actually observed, thus the selection effect is a matter only of the object size (apparent brightness).

The observations were performed with the Wise Observatory 0.46-m Centurion (C18). An SBIG ST-10XME CCD was used at the f/2.8 prime focus. This CCD covers a field of view of 40.5x27.3 arcmin with 2184x1472 pixels and was used in white light. Additional images were obtained with the Wise Observatory 1-m Ritchey-Chrétien telescope equipped with a cryogenically-cooled SITe CCD. At the f/7 focus of the telescope the CCD covers a field of view of 34x17 arcmin with 4096x2048 pixels. This telescope is equipped with filters that were used to determine the Atens colors.

The images were reduced in a standard way using bias and normalized flatfield images. We used the IRAF *phot* function for the photometric measurements. The photometric values were calibrated to a differential magnitude level, using 8-20 local comparison stars, and were afterwards calibrated using Landolt standards (Landolt 1992). In addition, the Aten magnitudes were corrected for light travel time and were reduced to a 1 AU distance from the Sun and the Earth to yield $H(1,\alpha^0)$ values (Bowell et al. 1989).

To retrieve the lightcurve frequency and amplitude, data analysis included folding all the calibrated magnitudes to one rotation phase, at zero phase angle, using two basic relations: the Fourier series for determining the variability period(s) (Harris and Lupishko 1989) and the H-G system for calibrating the phase angle influence on the magnitude (Bowel et al. 1989). The best match was chosen by least squares. The rotation period was deduced from the lightcurve and the object triaxial shape (assuming axes a ≥ b ≥ c) was calculated from the amplitude ΔM using:

$$\Delta M = 2.5 \cdot \log\left(\frac{a}{b}\right)_{min}.$$

**Research results**

The lightcurves of five Atens exhibit two types of features not usually seen in asteroids lightcurves: *i)* the presence of two or three independent and additive periods, and *ii)* the presence of frequencies with very high amplitude. In both cases, the assumption of binarity can be supported on different grounds. The other three Atens show single-body lightcurves.

As mentioned above, the first case can be interpreted as indicating the presence of an asynchronous binary where the rotational period of the primary is not synchronized with the orbital period of the secondary, therefore both periods are superposed on the lightcurve. An alternative interpretation for the different frequencies as a "tumbling" movement around a non-principal axis is less likely, due to the fast rotation of known asynchronous binaries compared to the slow rotation of known tumblers. In addition, the lightcurve of a tumbler asteroid is generally non-periodic with some characteristic frequencies present, and does not consist of only two additive frequencies as seen in the lightcurves of asynchronous binaries, but can be better described by a two-dimensional Fourier series (Pravec et al. 2005).

Two of the Atens, *1999 HF$_1$* and *2000 PJ$_5$*, show lightcurves with asynchronous binary characteristics. Both have fast rotating primaries (2.3192±0.0002 and 2.642±0.001 hours, respectively), and slow orbiting secondaries (14.017±0.004 and 9.446±0.004 hours). We should mention here that, while the binarity of *1999 HF$_1$* was first discovered by Pravec et al. (2002), *2000 PJ$_5$* was characterized as an asynchronous binary in our study. Fig. 1 shows the lightcurve example of *2000 PJ$_5$* obtained on July 9, 2005. The observed data are shown with the fitted model for the primary spin period and the residuals that represent a part of the orbital period. The orbital period curve clearly exhibits the presence of an eclipse produced by one of the components. The eclipse amplitude and duration can provide precise values for the component sizes in case a total eclipse occurred, or a limit to the component size ratio in case of a partial eclipse. From here, the Ds/Dp ratio can be calculated: this is 0.23±0.03 for *1999 HF$_1$* and 0.53±0.03 for *2000 PJ$_5$*.

The assumption of synchronous binaries can explain the presence of lightcurves with high amplitudes. Here, all three periods (primary and secondary spins and secondary orbital period) are identical and superposition can reveal the existence of three different motions only if they have very high amplitudes (>1 mag). This interpretation must take into account the possibility of a very elongated asteroid such as (216) *Kleopatra* or (1620) *Geographos*. One way to reject this option is the presence of steep V-shape minima in the lightcurve; these represent eclipse events, while wide U-shape maxima are produced by both components.

*1999 JD$_6$*, *2000 CK$_{33}$* and *2003 NZ$_6$* show these kinds of features in their lightcurves. Their amplitudes range from 1.16±0.05 mag (*1999 JD$_6$*) to 1.6±0.05 mag (*2003 NZ$_6$*) that correspond, respectively, for the case of elongated objects, to high a/b ratios of 2.91±0.05 to 4.37±0.05. In addition, the V-shaped minima of the lightcurves of *1999 JD$_6$* and *2000 CK$_{33}$* indicate probably mutual eclipses. The lightcurve of *2000 CK$_{33}$* is especially interesting (see Fig. 2) due to the asymmetry of one of its peaks. We measured a skew value of 0.64, higher than the expected value of 0.5. This asymmetry could be caused by an object with a crescent shape that causes the asteroid to cast a shadow onto its own surface at certain rotation phases. This could also be the signature of a very nearby or contact satellite rotating synchronously around the main body.

Using Kepler's III law, the determination of the binary system semi-major axis is possible by assuming a density for each object and by measuring the orbital period. We obtained a possible semi-major axis of 1.1 (*2000 CK$_{33}$*), 1.6 (*1999 JD$_6$*) and 2.3 (*2003 NZ$_6$*) Dp for these three synchronous binaries.

**Conclusions**

The high percentage of binaries and possible binaries reported here (five out of eight), suggests that binarity is a very common phenomenon for Atens, perhaps even more so for them than for other NEA groups, among which 15% are assumed to be binaries (Bottke and Melosh 1996). This may indicate that Atens suffer more disruptions caused by planetary encounters than do Apollos or Amors. A possible explanation could be their higher likelihood of tidal interactions with Earth and Venus, the massive terrestrial planets, which implies stronger tidal forces. Another option is the much closer distance of Atens to the Sun enhancing the effect of thermal re-emission due to YORP. As a result, Aten asteroids can rotate faster and could eventually disrupt. Pravec and Harris (2000) estimated that half of the fast-rotating NEAs are binaries, a claim that is consistent with our results (from a list of four fast-rotators, two are binaries). Whether this is true for slow NEA rotators as well, as our studies seem to indicate, should be further examined.

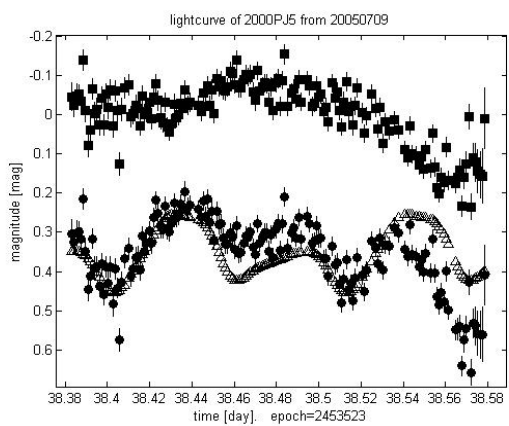

Fig 1: Lightcurve of *2000 PJ$_5$* from July 9, 2005; this is an example of an asynchronous binary lightcurve. Circles represent the observed data from which 18 mag were subtracted for display (note the three different peaks). Triangles show the fitted model with $P_1$=2.642±0.001 hours (primary rotation period), and squares show the residuals with a second period of $P_2$=9.446±0.004 hours (secondary orbital period). The start of an eclipse event can be seen on the top right.

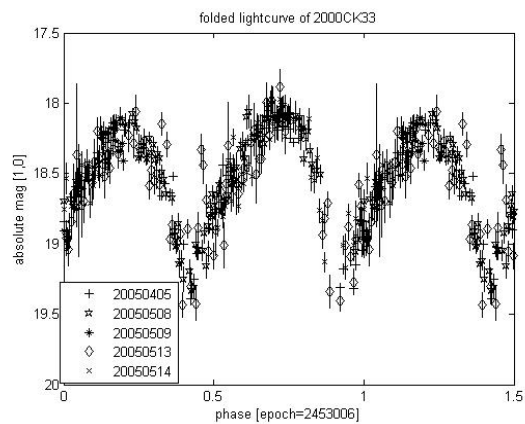

Fig 2: *2000 CK$_{33}$* lightcurve, folded with a period of 6.6045 hours. Notice the high amplitude, V-shaped minima and asymmetric peak (at phases ~0.5 to 1).

The existence of binaries within the Atens family is of major importance when considering any mitigation plan. The pulverization to dust of a binary asteroid by an immense explosion should consider the secondary orbit and location while detonating the explosive package, to avoid the shielding of energy by one of the fragments. In addition, a high efficiency could be achieved when the secondary is at periapsis. An explosion at the wrong orientation could divert the secondary possibly causing it to collide with the primary body with unpredicted consequences. However, one should recall that asteroid binarity may be a sign for the presence of a rubble pile structure. Hence, an explosion near a binary asteroid may result in many 100-200 meter-wide fragments that, in case of a short time to impact, may still poses considerable destruction power for the Earth.

When time to impact is not a factor, the asteroid deflection to a non-dangerous orbit, should be consider. Here as well, asteroid binarity should be considered. Nearby smaller explosions to deflect the asteroid into a different orbit, might result in crashing it into its companion. Gravitational dragging that involves near-"docking" with a tug close to the asteroid surface, would be reduced in efficiency by the presence of the satellite's gravity. A neutral location, at the center of mass of the system or above it but perpendicular to the orbital plane, could be too distant to produce a significant drag force. Painting or covering the asteroid surface with a specific albedo material will cause the thermal force to be a function of the fragment size only; this could separate between the primary and the much smaller secondary with unpredictable results.

A final caveat is that where there are two obvious companions to an NEA, there might well be more than two. One should consider NEAs as potentially multiple systems. The study of their physical properties is imperative to understand and evaluate the danger these bodies present.

**Summary**

Our photometric survey of NEAs from the Aten family found candidates for 2 asynchronous binaries and 3 synchronous binaries among 8 observed asteroids. This suggests that fraction of binaries, at least within the Atens, is higher than expected. Theoretical mitigation techniques should consider the possibility of binarity or multiplicity as a major characteristic of a potential and dangerous impactor.

**Acknowledgements**
Asteroid observations and studies in Israel are possible due to a grant from the Israel Space Agency and the Ministry of Science and Technology.

**Reference**
Bottke, W. F. and Melosh, H. J., 1996. Formation of asteroid satellites and doublet craters by planetary tidal forces. *Nature* **281**, 51-53.
Bottke, W. F. et al., 2002. The effect of Yakrovsky thermal forces on the dynamical evolution of asteroids and meteoroids. In: Bottke Jr., W.F., Cellino, A., Paolicchi, P., Binzel, R.P. (Eds.), Asteroids III. Univ. of Arizona Press, Tucson, pp. 395–408.
Bowell, E. et al., 1989. Application of photometric models to asteroids. In *Asteroids II* (R. P. Binzel, T. Gehrels and M. S. Matthews, Eds.), pp. 524-556. Univ. of Arizona Press, Tucson.
Harris, A. W. and Lupishko, D. F., 1989. Photometric lightcurve observations and reduction techniques. In *Asteroids II* (R. P. Binzel, T. Gehrels and M. S. Matthews, Eds.), pp. 39-53. Univ. of Arizona Press, Tucson.
Landolt, A., 1992. *UBVRI* Photometric Standard Stars in the Magnitude Range of $11.5 < V < 16.0$ Around the Celestial Equator. *The Astronomical Journal* **104**, no. 1, 340-371.
Merline, W. J. et al., 2002. Asteroids *do* have satellites. In *Asteroids III* (W. F. Bottke Jr. et al., eds.), Univ. of Arizona, Tucson.
Polishook, D. and Brosch, N., 2006. Photometry of Aten asteroids – more than a handful of binaries. *Icarus*, in press.
Pravec, P. and Harris, A. W., 2000. Fast and slow rotation of asteroids. *Icarus,* **148***,* 12–20.
Pravec, P. et al., 2002. Two periods of *1999 HF$_1$* - Another binary NEA candidate. *Icarus* **158**, 276-280.
Pravec, P. et al., 2005. Tumbling asteroids. *Icarus* **173**, 108-131.
Pravec, P. et al., 2006. Photometric survey of binary near-Earth asteroids. *Icarus* **181**, 63-93.